\documentclass[conference]{IEEEtran}
\IEEEoverridecommandlockouts
\usepackage{cite}
\usepackage{amsmath,amssymb,amsfonts}
\usepackage{algorithmic}
\usepackage{graphicx}
\usepackage{textcomp}
\usepackage{xcolor}
\usepackage{caption}
\usepackage{tcolorbox}

\usepackage{placeins}
\usepackage{float}
\usepackage{cleveref}
\usepackage{xcolor}
\usepackage{listings}

\lstdefinestyle{yaml}{
     basicstyle=\color{blue}\fontsize{6pt}{6pt}\selectfont,
     rulecolor=\color{black},
     string=[s]{'}{'},
     stringstyle=\color{blue},
     comment=[l]{:},
     commentstyle=\color{black},
     morecomment=[l]{-}
 }

\newcommand{\hide}[1]{Redacted for review}

\def\BibTeX{{\rm B\kern-.05em{\sc i\kern-.025em b}\kern-.08em
    T\kern-.1667em\lower.7ex\hbox{E}\kern-.125emX}}
\begin{document}
\title{PRE-Share Data: Assistance Tool for Resource-aware Designing of Data-sharing Pipelines}

\author{\IEEEauthorblockN{1\textsuperscript{st} Sepideh Masoudi}
\IEEEauthorblockA{\textit{Information Systems Engineering} \\
\textit{Technische Universität Berlin}\\
Berlin, Germany \\
smi@ise.tu-berlin.de}
}
\maketitle

\begin{abstract}
Data is a valuable asset, and sharing it as a product across organizations is key to building comprehensive and useful insights in fields such as science and industry. Before sharing, data often requires transformation to comply with governance policies and meet the requirements of recipient organizations. By leveraging pipelines, these transformations can be modeled as chains of processes; however, designing such pipelines while ensuring their efficiency is complex. In this paper, we present a tool that supports the design of pipelines by identifying opportunities for reusing transformation processes across different pipelines and suggesting designs and configurations based on these opportunities. This tool also generates reports on the resource consumption of pipeline processes, enabling the estimation of potential resource savings achievable through reuse-based designs. It could serve as a foundation for more efficient and resource-conscious data transformation pipeline design and be used as a component in self-service data platforms.

\end{abstract}

\begin{tcolorbox}[colback=white, colframe=black, boxrule=0.5mm, arc=0mm]
\paragraph*{Copyright Notice}
\textbf{Preprint}- This work has been accepted to the 22nd IEEE International Conference on Software Architecture Companion 2025(ICSA-C). Copyright may be transferred without notice, after which this version may no longer be accessible.
\end{tcolorbox}
\section{Introduction}Data is a critical asset that serves various purposes, from driving new revenue and capabilities in business to establishing a paradigm in science, alongside experimental, theoretical, and simulation-based approaches, to create new hypotheses and scientific discoveries\cite{grossman2023ten}.
Most organizations have already recognized that big data is essential for success and, consequently, have begun collecting data from multiple, diverse sources\cite{munappy2020data}.
However, transferring data in its original format and level of detail from the source is not feasible due to data governance requirements, policy enforcement, the volume, velocity, and variety of data across diverse sources, and the specific requirements of consumers.

With the increasing demand for data-sharing across organizations, concepts such as federated data products are gaining prominence. A data product is defined as a self-contained architectural quantum that integrates all the essential components required for its operation. This includes data and metadata, code for ingestion, processing, serving, and governance, as well as infrastructure for development, deployment, and storage\cite{goedegebuure2024data,datamesh}.
These involve dedicated teams responsible for managing the entire data product lifecycle, from ingestion to publication across multiple parties\cite{plebani2023teadal}.

Adopting product thinking to treat data as a interoperable product that meets the needs of data consumers and can be seamlessly combined and transformed to achieve a greater higher-order value, led to emergence of data platforms to enable organizations to create data products with hight autonomy\cite{van2024architectural}. These platform which are called self-serve data platform\cite{datamesh}, aim to lower the barriers for domain teams in to own, build, and exchange data products and for governance teams to monitor and ensure interoperability, compliance, and quality of data products\cite{van2024architectural}. 

When building a self-serve data platform, a platform team (e.g., platform architects and engineers) needs to consider various architectural design decisions and select the appropriate options for them to optimize the experience for data product teams, consumers, and governance teams\cite{van2024architectural,ashraf2023key,datamesh}.

Data products are typically designed to address the needs of a target group of consumers. However, the data within these products often requires further transformation to cater to the unique requirements of individual consumers, resulting in what are referred to as consumer-aligned data products \cite{falconi2024data,datamesh}.
Data-sharing pipelines, as a part of infrastructure utility plane of self-service data platforms for executing data product components\cite{van2024architectural}, are a solution for seamlessly sharing the data of a data product through a complex chain of interconnected processes aimed at transforming data to meet the requirements of a specific consumer, as well as data governance and policy restrictions.
However, data transformation pipeline design, development, management, and maintenance are complicated tasks that require considering a broad range of design options and selecting between them\cite{van2024architectural}.
These design decisions include approaches for customizing and reusing pipelines, defining pipeline templates, connectors, and orchestrations, as well as strategies for monitoring pipelines to dynamically enhance the system as new pipelines are added to support new consumers in using the data products\cite{van2024architectural,wider2023decentralized}.
Therefore, by supporting common types of data transformation processes defined as parameterized templates, the data platform enables the development of various pipelines, leading to customized data products for different consumers.
Moreover, strategies for enabling the reuse of pipelines and transformation processes across pipelines is one of the major design decisions that should be tackled in order to support customization of pipelines in a more optimized manner during the data product lifetime\cite{van2024architectural}.
Most companies handle this maintenance manually by appointing a dedicated person to administer the data flow through the pipelines\cite{munappy2020data}. 
Though, with the increasing number of unique customized pipelines between a data product and each of its consumers, it becomes increasingly difficult to find common processes and reuse-based design options across different pipelines, especially if we are aiming to leverage reuse for improving a specific quality, such as resource consumption and energy consumption.

Since a data transformation pipeline is a connected chain of processes in which the output of one process becomes the input for another\cite{raman2013beyond}, designing these pipelines through reusing a subset of this chain across multiple pipelines can create new opportunities to reduce both effort and resource consumption.
However, the possibilities and complexities involved in designing data transformation pipelines and identifying opportunities for reuse highlight the need for tools that support intentional design choices. These tools enable data-sharing producers to create pipelines that are more reusable and resource-efficient, leveraging highly parameterized, predefined transformation process templates.
In this paper, we introduce a cloud-based tool aimed at automating the discovery of reuse opportunities and providing the capability to configure and estimate the impact of those reuse strategies on the resource consumption of data transformation pipelines by providing insight into the resource consumption of every process that is part of the pipeline in each configuration of them.

The remainder of this paper is organized as follows: In the next section, we review related work, focusing on data platforms and transformation pipelines in data products. In Section 3, we introduce our proposed tool for discovering reuse opportunities and providing an estimation report on its effect on resource consumption. In Section 4, we describe the application of this tool and its implementation, and in Section 5, we conclude with a discussion of future work.\label{introduction}
\section{Related Work}Van et al. study the various planes of a self-service data platform that provides services for building, sharing, and managing data products efficiently \cite{van2024architectural}. They also define and categorize key design decisions to consider when implementing and deploying a self-service data platform, including selecting an approach for reusing pipelines and utilizing pipeline templates, a challenge our tool aims to address.

Papadakis et al. present a software overlay for addressing the challenges of data sharing between heterogeneous federated data spaces by improving modularity, scalability, and interoperability\cite{papadakis2024ccduit}.
Additionally, Grossman defines data commons as cloud-based data platforms with a governance structure that allows a community to manage, analyze, and share data\cite{grossman2023ten}. 
The author also mentions cost reduction as one of the main purposes of using data commons. In this study, we aim to address this aspect by developing a tool that seeks to reduce resource costs by promoting reuse in data transformation pipelines. Thus, in our paper, we provide a tool to develop an understanding regarding the reuse strategies and their impact on resource consumption in these studies.

Hofman proposes a federated platform service for sharing data across multiple global parties, using pipelines to model data collection and transformation between parties\cite{hofman2015towards}. 
Dehury et al. present an extension of the Tosca standard, Toscadata, which focuses on modeling data pipeline-based cloud applications. The authors leverage TOSCA models and a serverless platform to manage the flow and transformation of data in a pipeline structure\cite{dehury2022toscadata}. 
This study also seeks to increase the efficiency of data pipelines by implementing them as cloud applications that are independently scalable and deployable. However, these studies do not provide a mechanism for discovering potential reuse strategies with focus on pipelines.

Stutterheim et al. develop a distributed data-sharing platform utilizing a composable chain of microservices to shape the data exchange structure. They propose using a knowledge base repository where all microservices are labeled and stored, and a central orchestrator uses those microservices to build a chain of processes required to handle data based on user requests\cite{stutterheim2024dynamos}. However, they did not investigate the reuse of deployed microservices in different chains or building a chain while considering its common processes with other chains.

Consequently, our paper presents an implemented tool that facilitates and automates reuse-based pipeline design and generates resource consumption reports to support more informed design decision-making.

\label{relatedworks}
\section{Architecture}The self-service data platform comprises different planes, including the infrastructure plane, the data product experience plane, and the mesh experience plane, which aim to standardize and improve the experience of developing and utilizing cross-organizational data products for both the maintenance team and data consumers\cite{van2024architectural,falconi2023adopting}.
Transforming data is one of the key capabilities that should be offered by the infrastructure plane in a self-service data platform. This capability can be modeled as pipelines to facilitate the orchestration of the transformation process required to adjust data to meet consumer requirements\cite{munappy2020data,van2024architectural}.

However, designing data transformation pipelines involves considering various criteria, including data governance and policies, compliance with agreements between data providers and consumers, and other relevant aspects. Notably, recent studies emphasize the importance of creating resource-conscious pipelines\cite{plebani2023teadal}. Moreover, finding an approach to support customization and especially reuse of pipelines is one of the design decisions that should be addressed\cite{van2024architectural}.
Therefore, facilitating the design of pipelines by discovering options for reusing common processes across different pipelines or configuring pipelines to serve multiple data consumers, while providing insights into the resource consumption of processes within the pipelines, establishes a foundation for enabling reuse-based design decisions with resource awareness.

\begin{figure}[htbp]
\centerline{\includegraphics[height=0.3\textheight, width=0.3\textwidth]{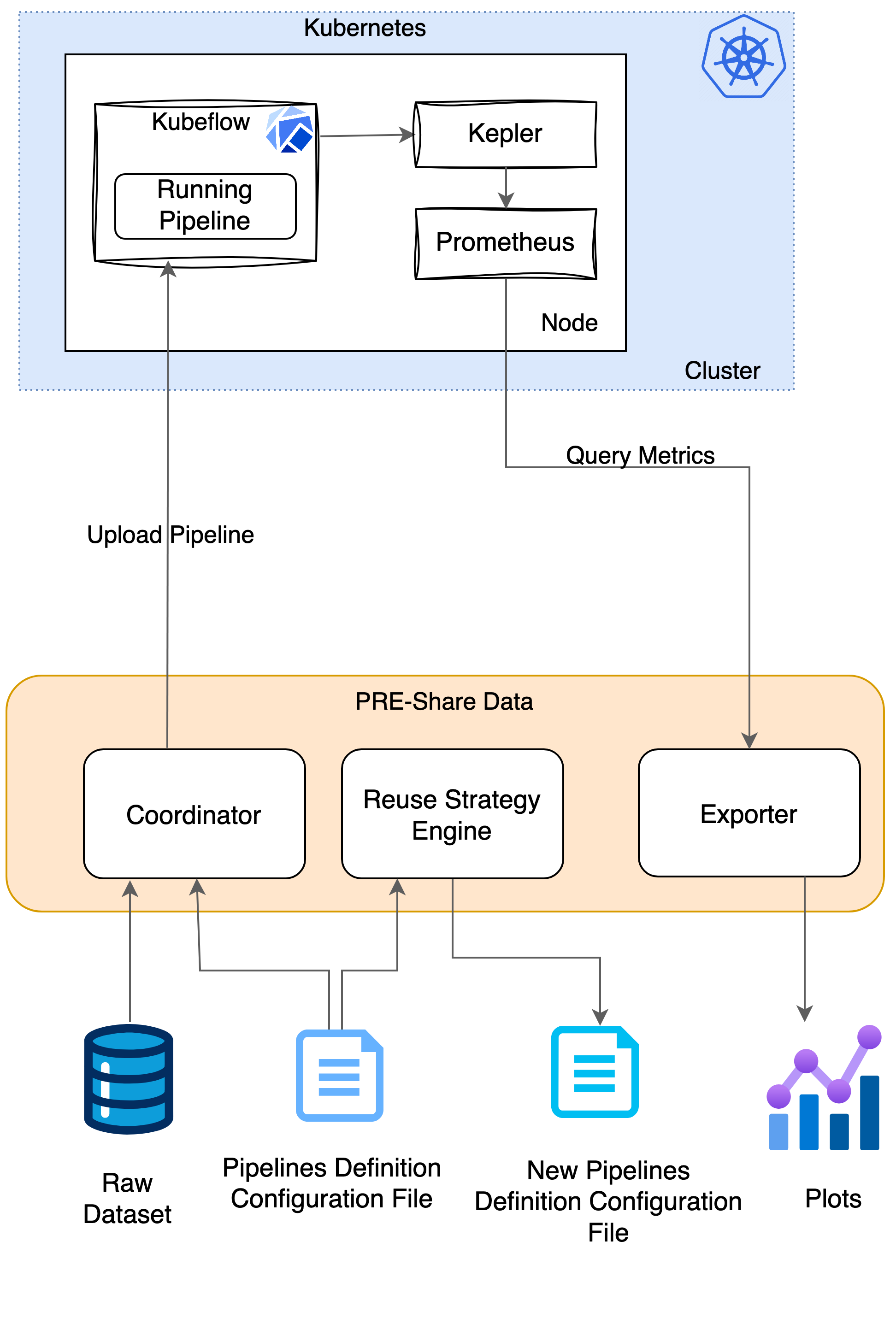}}
\caption{Architecture of the PRE-share Data tool: an assistance tool for discovering resource-aware reuse-based designs in pipeline configurations.}\label{arch}
\end{figure}

We propose \textit{PRE (\textbf{P}ipeline \textbf{Re}use) -share Data}, a tool to assist in the design and configuration of pipelines in a data-sharing context by detecting reuse opportunities across data transformation pipelines originating from the same data product. This tool is open source and available on GitHub\footnote{GitHub address: https://github.com/Sepide-Masoudi/kubeflow-data-sharing-pipeline}.
As shown in \Cref{arch}, this tool builds, uploads and runs transformation pipelines on the Kubeflow platform\footnote{https://www.kubeflow.org/} based on a user-provided configuration file, facilitated by the Coordinator component.
Currently, PRE supports filtering, aggregation, anonymization, and formatting as embedded transformation processes that serve as templates for building pipelines. These templates allow users to utilize them by providing desired parameters and arguments in the configuration file. However, the tool is easily extendable to support other customized processes as part of a pipeline.
Next, the Reuse Strategy Engine iterates over the definition of pipelines to discover potential process reuse across pipelines based on the shared named process.
This module generates a new execution configuration file that reflects reuse strategies across data transformation pipelines. The new configuration file contains pipelines that include only the common processes shared across consumers, as well as pipelines that contain unique processes for each consumer. In the new configuration, there is a chain of pipelines that reuses the pipeline with common processes across all pipelines, alongside pipelines that contain unique processes required for specific consumers, ensuring the proper sequence and creating the new reuse-based configuration for running pipelines.
Additionally, the Exporter component generates reports by gathering metrics from Prometheus, detailing the resource consumption of each process in the pipeline and the total resource consumption of the pipeline execution. This information provides the pipeline designer with an estimate of the resource savings achievable through pipeline redesign and reuse of common processes. Furthermore, the new configuration file generated by the Reuse Strategy Engine is also runnable by the tool, allowing the designer to use this configuration to measure the resource consumption of the alternative configuration precisely.
It is worth mentioning that the pipeline designer or system administrator is responsible for deploying Kepler and Prometheus in the operational environment. Our tool then automates the deployment and execution of pipeline based on the input configuration file, generating a new configuration file that reflects possible reuse and reports about the resource consumption of data transformation pipelines. 

\label{energyProfiler}
\section{Implementation And Demonstration}For describing the Procedure of \textit{PRE-Share Data}, we present a use case that demonstrates its utility in configuring and designing data transformation pipelines. In this scenario, an open source industrial dataset\footnote{Dataset address: http://industry.teadal.ubiwhere.com/fdp-czech-plant/shipments}belongs to TEADAL\footnote{https://teadal.eu/} project, serves as the data product intended for sharing with consumers, and its raw data is accessible via a REST URL.
We have four options for transformation processes, including filtering, aggregation, anonymization, and formatting in JSON and CSV formats. Although the tool is not limited to these processes, the user can easily replace them with any alternative processes simply by adding a python function or providing the Docker image addresses of the processes.
To initiate the tool's operation, the user needs to upload a YAML configuration file in a specific format. As shown in \Cref{config}, this file specifies the required processes, the necessary arguments, and values for running these processes. It also includes pipeline definitions that indicate the sequence and structure of processes within each pipeline.  
Subsequently, \textit{PRE-Share Data} reads this configuration file, builds Kubeflow components for each process, defines the pipelines by integrating these components, and uploads the pipelines to Kubeflow. Once the pipelines are running on the Kubeflow platform, Kepler measures the metrics for each process and pipeline. Upon completion, \textit{PRE-Share Data} generates plots showing resource consumption for each process and the overall resource consumption of the pipelines.
Moreover, \textit{PRE-Share Data} provides an alternative configuration file based on the original. This new configuration file, as shown in \Cref{reuse-config}, is built by reusing common processes across different pipelines. The alternative configuration is also compatible with \textit{PRE-Share Data}, allowing the pipeline designer to rerun the pipelines with this new setup to assess the impact of reuse-based design on resource consumption. However, even without executing the new configuration, the system designer can use the generated plots to estimate potential resource savings from reuse.
It is worth mentioning that, while the configuration file format is specific to \textit{PRE-Share Data}, the tool also produces a deployment file compatible with Kubeflow. This file is saved in the root directory so that, after finalizing the design phase, the final configuration output can also be directly used in the Kubeflow platform for deployment in the production environment.

    \begin{figure}[htbp]
    \centerline{}
        \begin{lstlisting}[style=yaml]
    Deployment:
      namespace: kubeflow
      prometheusURL: 'http://localhost:9090'
    stages:
      - name: filter_data
        type: filtering
        parameter:
          operation: 'greater_than'
          column_name: 'week'
          threshold: 3
      - <other stages ...>
    pipelines:
      - name: shipments_anonymize_data_pipelines
        flow:
          - filter_data
          - anonymize_columns
          - aggregate_data
          - compress_json_to_output
        consumers:
          - u1
      - name: shipments_data_pipelines
        flow:
          - filter_data
          - aggregate_data
          - compress_json_to_output
        consumers:
          - u2
    pipelineChains:
      - name: shipments_anonymize_chain
        flow:
          - shipments_anonymize_data_pipelines
    
      - name: shipments_data_chain
        flow:
          - shipments_data_pipelines
        \end{lstlisting}
        \caption{Sample input YAML configuration file for PRE, defining the pipelines and their parameter values for execution.}
        \label{config}
    \end{figure}
  
\begin{figure}[htbp]
\centerline{}
    \begin{lstlisting}[style=yaml]
    <same as input YAML...>    
pipelines:
- name: auto_gen_pipeline_b5e81b9e-6e74-48e9-aa1a-4b5a66bf3109
  flow:
  - filter_data
  consumers:
  - u1
  - u2
- name: auto_gen_pipeline_05ec9b5a-3eda-49e8-835e-fc25412e170a
  flow:
  - anonymize_columns
  consumers:
  - u1

- name: auto_gen_pipeline_b6b47918-85a3-4060-95e2-821bf074333c
  flow:
  - aggregate_data
  - compress_json_to_output
  consumers:
  - u1
  - u2
pipelineChains:
- name: shipments_anonymize_data_pipelines_00b3aef4-a0c9
  flow:
  - auto_gen_pipeline_b5e81b9e-6e74-48e9-aa1a-4b5a66bf3109
  - auto_gen_pipeline_05ec9b5a-3eda-49e8-835e-fc25412e170a
  - auto_gen_pipeline_b6b47918-85a3-4060-95e2-821bf074333c
- name: shipments_data_pipelines_2fb7a6ed-45ce-42c7-8649-8e13ac1d3439
  flow:
  - auto_gen_pipeline_b5e81b9e-6e74-48e9-aa1a-4b5a66bf3109
  - auto_gen_pipeline_b6b47918-85a3-4060-95e2-821bf074333c
    \end{lstlisting}
    \caption{Sample reuse-based configuration generated by PRE, formatted for execution in PRE. PRE will define new pipelines of common process and create chain of pipelines by reusing new pipelines across different pipelines. Auto generated pipelines names start with 'auto-gen.*'.}
    \label{reuse-config}
\end{figure}
\label{evaluation}
\section{Conclusion}In this paper, we present \textit{PRE-Share Data}, a tool that provides insights into potential reuse strategies for configuring and designing data transformation pipelines of a data product, as well as estimating the resource savings achievable through reuse. Our tool assists in designing pipelines by offering alternative, reuse-based configuration options while utilizing Kubeflow as the platform and can be used in data platforms. In the future, we plan to enhance the adaptability of PRE by making it compatible with additional pipeline platforms beyond Kubeflow, thus increasing its broader applicability. We also aim to enrich its reporting capabilities to cover cost and performance metrics, further supporting informed decisions in pipeline configuration.
\label{conclusion}

\section*{Acknowledgment}
Funded by the European Union (TEADAL, 101070186). Views and opinions expressed are however those of the author(s) only and do not necessarily reflect those of the European Union. Neither the European Union nor the granting authority can be held responsible for them.

\bibliographystyle{IEEEtran}

\end{document}